# Intrapixel Effects of CCD and CMOS Detectors

**Hu Zhan[*], Xin Zhang and Li Cao**

*CAS Key Laboratory of Space Astronomy and Technology, National Astronomical Observatories,*
*A20 Datun Road, Chaoyang District, Beijing 100012, China*
*E-mail*: `zhanhu@nao.cas.cn`

ABSTRACT: Intrapixel nonuniformity is known to exist in CCD and CMOS image sensors, though the effects in backside illuminated (BSI) CCDs are too small to be a concern for most astronomical observations. However, projects like the Large Synoptic Survey Telescope require precise knowledge of the detector characteristics, and intrapixel effects may need more attention. By scanning CCD and CMOS cameras with a small light spot (unresolved by the optics), we find in the images that the spot's flux, centroid displacement, and ellipticity vary periodically on the pixel scale in most cases. The amplitude of variation depends on not only the detector but also how well the spot is sampled by the pixels. With a spot radius of 2 pixels (encircling 80% energy) as measured, the flux and the ellipticity extracted from the BSI CCD camera vary by 0.2-0.3% (rms) and 0.005 (rms), respectively, while the deviation of the centroid position (rms ~ 0.01 pixel) is not correlated with the pixels. The effects are more pronounced for the BSI CMOS camera and even worse for the frontside illuminated CMOS camera. The results suggest that a closer examination of the intrapixel effects is needed for precision astronomy.

KEYWORDS: Photon detectors for UV, visible and IR photons (solid-state) ; Systematic effects; Image processing

---

[*] Corresponding author.

**Contents**

**1. Introduction**

The use of CCDs in optical astronomy has made it possible to collect a huge volume of high-quality data for precision measurements. For example, the Sloan Digital Sky Survey (SDSS) is able to achieve relative photometry errors of ~1-2% over roughly a quarter of the sky [1], which is an important factor for its great success. Demanding science goals of ongoing and future survey projects such as the GAIA mission [2] and the Large Synoptic Survey Telescope (LSST) [3] continue to pursue the limit of precision astronomy. The case of probing dark energy through minute weak lensing effects places stringent requirements on photometry, astrometry, and shape measurements. It is hence not surprising that detectors, being a critical component of the whole telescope system, have been under close examination for potential systematic effects on the measurements and subsequent cosmological studies [4][5][6].

Ideally, one expects the detector to faithfully record the incident photons. In the absence of statistical noise, pixelization can alter the estimated centroid of a star in a deterministic way but should not affect its flux. If we consider nonuniformity within each pixel, either because of the material or electric field inside, then the measured flux of an undersampled image also depends on the true position of the star's centroid within the pixel. Such an intrapixel effect is present not only in front-side illuminated (FSI) devices but, to a lesser degree, also in backside illuminated (BSI) devices [7][8][9][10]. However, as we show in this work, even in the case where sampling is reasonably good, e.g., with a radius encircling 80% energy ($R_{EE80}$) of the point spread function (PSF) of 2 pixels, intrapixel effects may still be relevant to projects like the LSST.

While CCDs remain the detector of choice for optical astronomy, CMOS detectors have made significant improvement on a number of performance characteristics such as quantum efficiency and readout noise [11]. Moreover, CMOS detectors' ability to read out quickly and address pixels individually is highly desirable for time domain observations. In fact, BSI CMOS detectors are used by the Transneptunian Automated Occultation Survey (TAOS II) [12]. The advantage of the CMOS detectors is a result of the active pixel architecture, i.e., each pixel physically containing its own amplifier(s), which, unfortunately, also gives rise to intrapixel photoresponse variations and interpixel cross talks [13].

Given the requirements of precision astronomy on CCDs and the need to evaluate CMOS detectors for astronomical observations, we carry out spot-scan tests on both types of detectors (cameras) to quantify the intrapixel effects under reasonably well-sampled conditions. Besides the spot image's flux and centroid position, we also perform tests on its ellipticity, which is a key quantity to measure for weak lensing studies.

## 2. Test setup

Our test setup is similar to that in reference [7]. As illustrated in Figure 1, a 10μm-diameter pinhole is imaged onto the detector by a commercial camera lens of focal length $f$=50mm. The pinhole and the light source are enclosed in a light-tight housing and mounted on a precision stage (X-stage) moving horizontally and, at the same time, perpendicular to the optical axis. The X-stage is then mounted on a vertical stage to allow adjustment in the other direction. The X-stage has an on-axis accuracy of 5μm. Since the distance between the pinhole and the lens is approximately 870mm, the motion of the spot image on the detector can be controlled to within 0.3μm, sufficient for scanning pixels of size ≳10μm. Moreover, the pinhole cannot be resolved at such a distance even if the aperture of the lens is open to the maximum (f/1.8). Hence, the spot image is essentially the PSF of the optics itself.

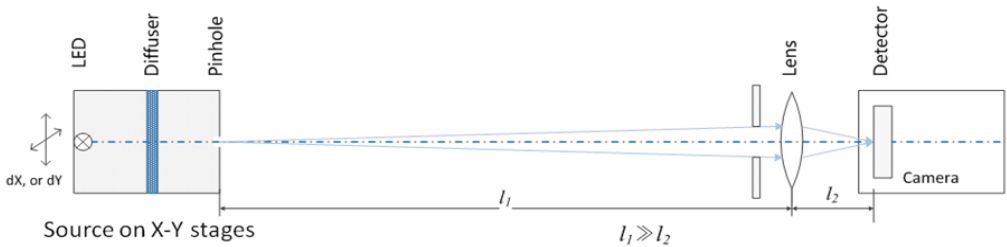

**Figure 1** Schematic diagram of the test setup.

The whole test system is assembled on a damped optical table. The test room maintains a constant temperature nominally, but there is no active thermal control on the apparatus. A 0.1℃ rise of the temperature would increase the length of a 200mm-long stainless steel rod by nearly 0.2μm. We in fact see gradual motions of the spot image of ≲0.2μm in the vertical direction and less in the horizontal direction even if the pinhole does not move. Since the thermal effect varies slowly and is uncorrelated with the pixel scale, it cannot mimic the intrapixel effects.

The detectors tested include an e2v CCD47-20 (in an Andor DV435 camera operating at -40℃) and a GSENSE400 BSI VIS CMOS image sensor from Gpixel Inc. (on an evaluation board at room temperature; we still call it a camera). The pixel size of the CCD is 13μm, and that of the CMOS detector is 11μm. The thickness of the silicon and operating voltages have a significant impact on the detector's performance. Unfortunately, we do not have such information about the CCD, but it is consistent with a 13μm thick standard silicon device operating in the inverted mode without a backside bias voltage (private communication with P. Jorden from e2v). The thickness of the silicon in the CMOS detector is 3.6μm, and the maximum voltage supplied is 3.3V.

Both cameras output 16-bit pixel values, though the CMOS detector combines on chip two 12-bit readout chains, one high-gain and the other low-gain, to obtain 16-bit results. Although the detector's dark current depends on the operating temperature, the intrapixel effects are unlikely to be overly sensitive to it. Nevertheless, we plan to build a cryogenically cooled camera to test the CMOS detector in the future. We have also examined three so-called scientific CMOS cameras from two vendors. These cameras all use FSI detectors and suffer much worse intrapixel effects.

We use a "blue" (365nm) LED and then a "red" (850nm) LED as the light source to see if the intrapixel effects depend on the wavelength. The LEDs are lit by a stable driver. We assess the stability of the LED's output light power by carrying out the spot scan without actually moving the pinhole. After warming up, the light power drifts roughly 0.1% with the red LED and less with the blue one over the course of a single scan (40 min). The spectra of the two LEDs are shown in Figure 2. Since there is no strong narrow line in the spectra, fringing is unlikely to obscure the test results. For convenience, we address the spot-scan results below as if they were taken at the nominal wavelength of the LEDs, even though the LEDs are by no means monochromatic.

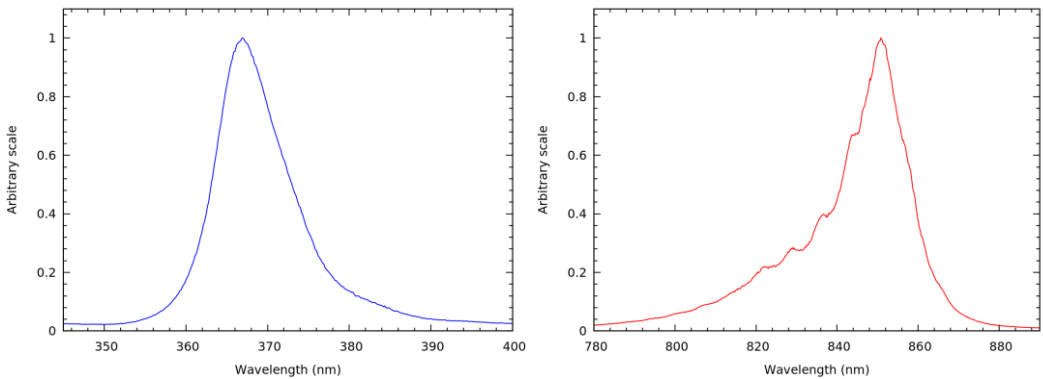

**Figure 2** Spectra of the 365nm LED (left, resolution 0.34nm) and the 850nm LED (right, resolution 0.29nm).

The aperture of the lens is set to f/11, and the image is adjusted slightly out of focus to produce a PSF with $R_{EE80}$ = 2 pixels, i.e., 26μm for the CCD and 22μm for the CMOS detector. The amount of defocusing is small enough so that the PSF can still be approximated by a Gaussian. The step size of the pinhole in the image plane is roughly 0.1 pixel. The exposure time is adjusted for each camera so that the brightest pixel is at roughly half of its full well capacity. Ten (sixteen) images are taken at each step with the CCD (CMOS) camera and averaged after bias subtraction and flat-fielding. The photon noise of the averaged spot image at each step is 0.08% or better for both cameras.

## 3. Intrapixel effects

We use SExtractor [14] to measure the spot image's flux (FLUX_APER, PHOT_APERTURES =10), centroid position (X_IMAGE and Y_IMAGE), and ellipticity (ELLIPTICITY). The images are convolved with a 3×3 mask ({{1,2,1},{2,4,2},{1,2,1}}) before being processed. While smoothing helps suppressing random fluctuations in the results, it also makes them less sensitive to nonuniformity inside the pixels. A relative detection threshold (DETECT_THRESH) of 50 has been applied. Lowering it to 5 slightly decreases variations of the centroid and ellipticity and has essentially no effect on fractional variations of the flux. Figure 3 shows the results extracted from simulated images of a circular Gaussian spot ($R_{EE80}$ = 2 pixels) moving uniformly along the center of a row of pixels. The precisions achieved on the flux, centroid position, and ellipticity measurements are 0.1% (photon noise limited), 0.001pixel, and 0.002, respectively. Results with pixel weighting (e.g., XWIN_IMAGE, etc.) would be less noisy, but the scheme assigns more weight to the central part of the spot image, which would amplify the intrapixel effects. Given the treatment of smoothing and adoption of the isophotal measurements from SExtractor, our results can be taken as conservative estimates of the intrapixel effects.

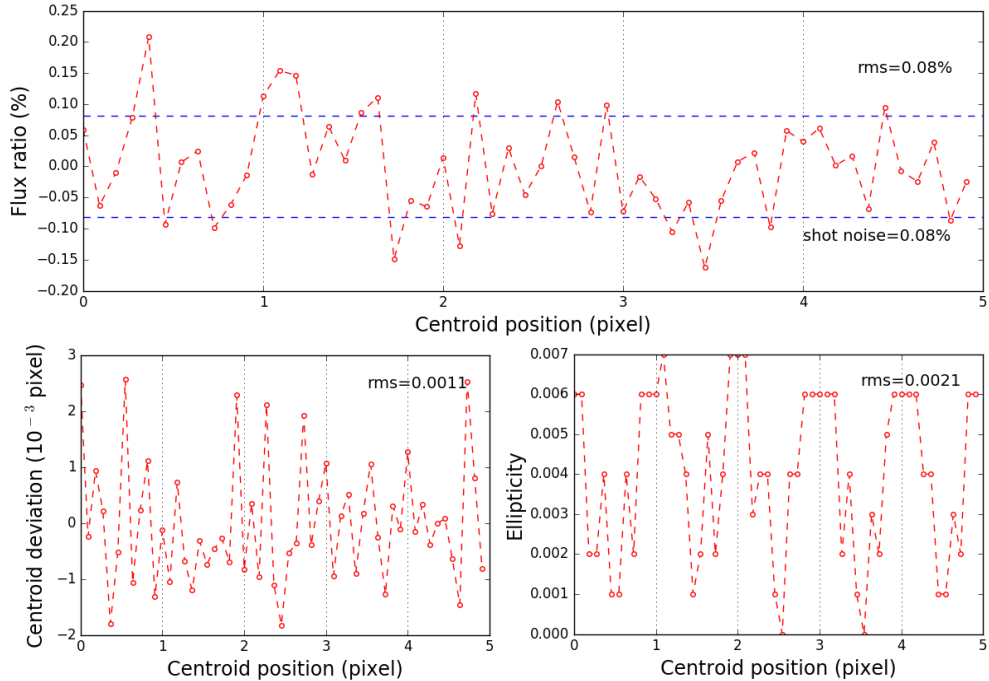

**Figure 3** Variations of the simulated spot's relative flux (top), centroid deviation from the true position along the direction of motion (lower left), and ellipticity (lower right). The flux is normalized by the mean value.

The variations of the real spot image's flux, its centroid relative to the X-stage's position in the image plane along the direction of motion, and its ellipticity as the spot image moves across pixel columns are presented in Figure 4, Figure 5, and Figure 6, respectively. One sees that both the BSI CCD and the BSI CMOS detectors show some degree of fluctuations on the pixel scale in at least one wavelength, except the inconclusive centroid test on the CCD. The results indicate that the variations originate from the interior of the pixel because other sources of influence have no knowledge of the pixel scale. Moreover, the CMOS detector displays larger variations than the CCD in nearly all cases, especially at 850nm.

### 3.1 Effect on flux measurement

The flux in Figure 4 is measured with an aperture of 10 pixels, and the value at each step is stable to better than 0.1% as the aperture increases further. The rms of the flux variations is 0.2-0.3% with the CCD camera and several times larger with the CMOS camera. The CCD result at 365nm is inconclusive because the variations are not correlated with the pixel scale. We notice that pronounced periodic patterns appear in the CCD's flat field images at 365nm, so the result in the upper left panel of Figure 4 could be affected by the residual of flat fielding. Similar behavior in the near ultraviolet is observed with other CCDs as well.

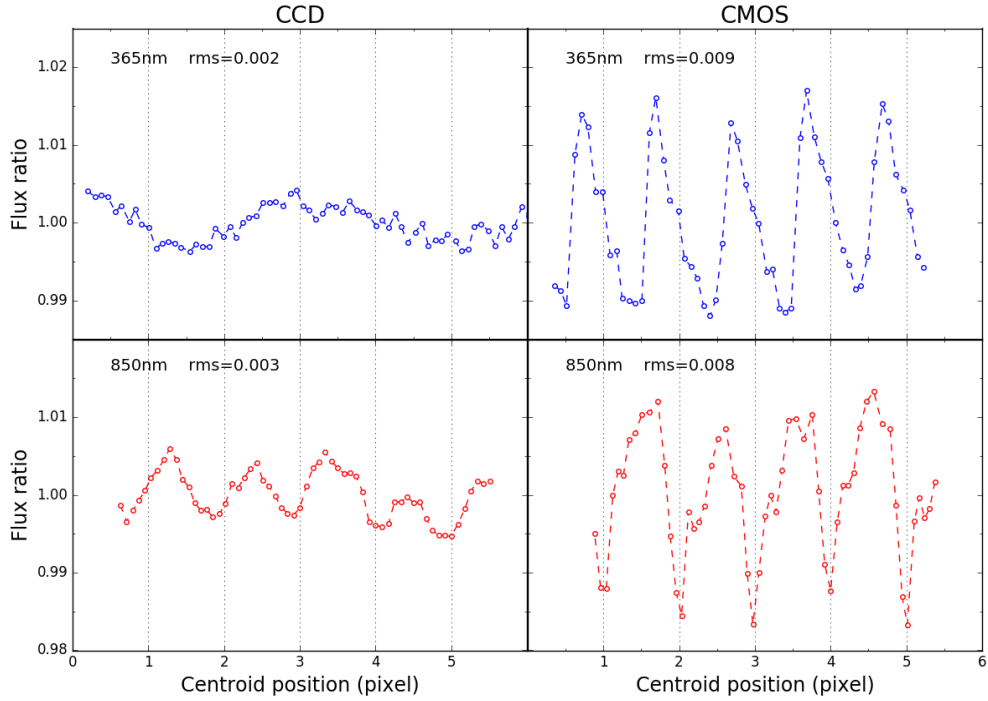

**Figure 4** Variations of the spot flux relative to its average as the image moves across pixel columns on the CCD (left panels) and CMOS (right panels) detectors. The data in the upper (lower) panels are taken under illumination of the 365nm (850nm) LED.

**3.2 Effect on centroid measurement**

We map the X-stage's motion onto the image plane as a reference to examine how nonuniformity inside the pixels affects the centroid position of the spot image. The 5μm on-axis accuracy of the X-stage corresponds to ~0.02 pixel in the image plane, marginally suitable for the centroid test. In practice, the X-stage can achieve better accuracy if it moves in short ranges, e.g., ≲1mm for the spot scans, and always approaches the same position in the same direction.

The CCD's centroid results in Figure 5 are again inconclusive, regardless which LED is used. The overall trend in the CCD data is likely due to small temperature variations rather than the inaccuracy of the X-stage. Even though there is a hint of variations on the pixel scale in the CMOS detector's centroid data at 365nm, the amplitude is not much larger than those of the CCD results. Therefore, we consider the CMOS result at 365nm inconclusive as well. However, pixel-scale variations in the CMOS result at 850nm can be seen clearly.

**3.3 Effect on ellipticity measurement**

Ellipticity measurement is crucial to weak lensing studies. Although it is not necessary to work with perfectly round spot images in this test, we still make sure that the lowest ellipticities in each panel of Figure 6 are as close to zero as possible. The CCD result at 365nm does not show an obvious correlation between the ellipticity and the centroid position for half of the data, but the cases in the other panels are clear. The amplitude of the variations is large compared to typical weak lensing signals and will cause systematic errors if uncorrected.

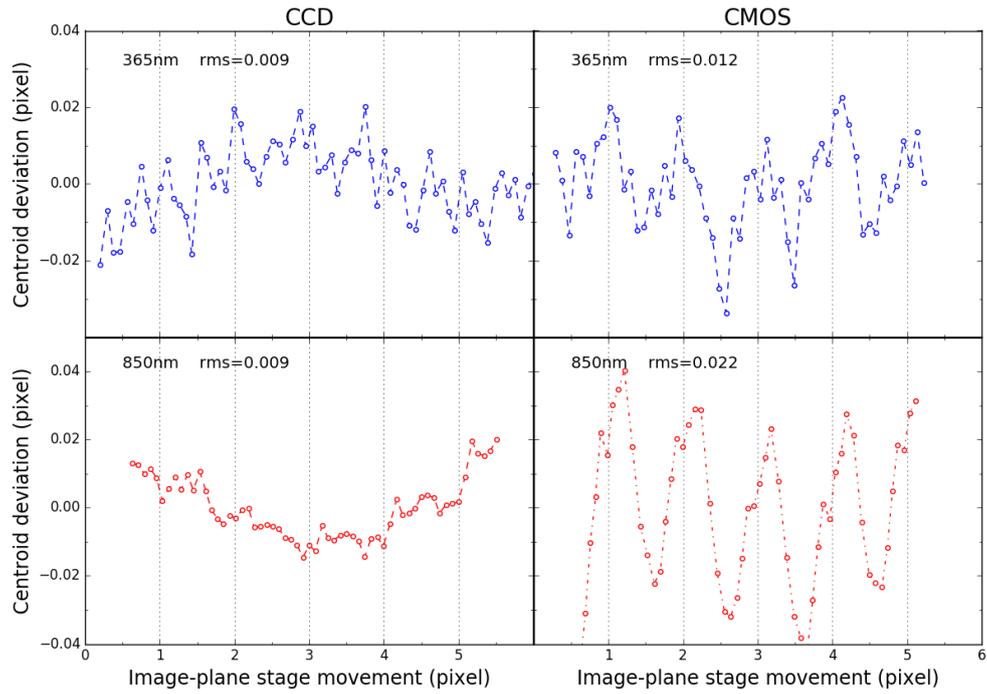

**Figure 5** Same as Figure 4 but for deviations of the spot centroid, in the direction of motion, from the image-plane position of the X-stage.

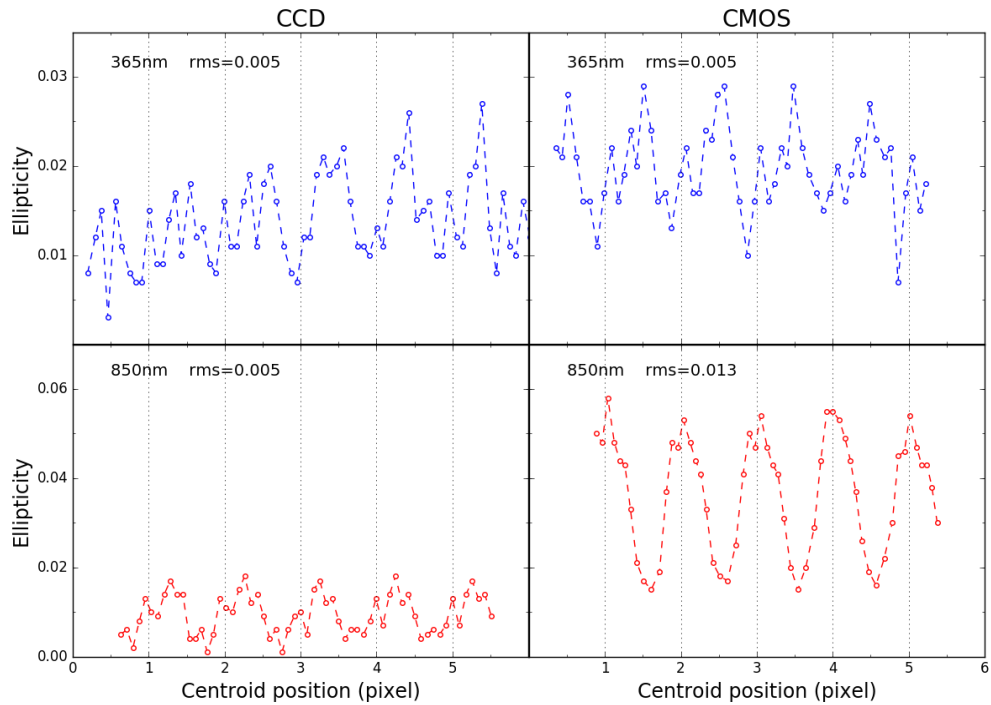

**Figure 6** Same as Figure 4 but for variations of the spot ellipticity.

## 4. Summary

We find through spot scans that non-negligible intrapixel effects exist in both BSI CCD and BSI CMOS detectors reported in this work. These effects are also seen in other CCD and CMOS detectors that we have tested, including a deep-depletion CCD. The 0.2-0.3% variations in the measured spot flux in the reasonably well sampled CCD images would seem to be a significant source of systematic errors if one wishes to achieve 0.5-1% relative photometry. Similarly, the spot's ellipticity variations of 0.005 in the CCD images are not negligible for weak lensing studies. The test of the centroid position is inconclusive with the CCD, but, if we take the 0.01 pixel rms error in the experiment as a limit, the induced astrometry error for a pixel size of 0.2arcsec would be 2mas, which is worse than the LSST goal. Hence, the intrapixel effects are likely to be relevant to surveys aiming for precision measurements.

It is interesting that the centroid and ellipticity results of the CMOS detector vary significantly more at 850nm than at 365nm. Optical aberrations are not likely to be the major factor, because the CCD results under the same test conditions are much less affected by the wavelength. Given that longer-wavelength photons are absorbed deeper in the silicon, a plausible explanation may lie in the thickness of the silicon in the CMOS detector, which is less than one-third of that in the CCD.

It is somewhat surprising that the total flux of the spot image on BSI detectors can depend on its centroid position within a pixel. Reference [7] attributes this effect in BSI CCDs to a possibly lower collection probability of electrons generated outside the depletion region. The pixel structure of BSI CMOS detectors is considerably more complex than that of CCDs, so we expect to see stronger intrapixel effects with BSI CMOS detectors. Nonuniformity inside the pixels and undersampling of the PSF together give rise to the intrapixel effects. To mitigate the problem, one can either increase the pixel sampling rate or, if the PSF is stable, dither on sub-pixel scales. Meanwhile, it is also worth developing algorithms to correct for the intrapixel effects.

Finally, we note that the intrapixel effects depend on the very design of the detector, the imaging system, and the operations. Therefore, our test results are not directly applicable to specific projects. To assess the impact on a particular project, one should evaluate the intrapixel effects under the project's realistic observational conditions.

## Acknowledgments

We thank J.A. Tyson, P. Jorden, C. Stubbs, and R. Lupton for discussions. This work was supported by the China Manned Space Program and the Astronomical Facility Renovation and Major Instrument Operations Fund from Chinese Academy of Sciences.